\documentclass[prl,twocolumn,aps,amsmath,amssymb]{revtex4}

\usepackage{graphicx}
\usepackage{dcolumn}
\usepackage{bm}
\usepackage{amsmath}
\usepackage{amsfonts}

\def\drawbox#1#2{\hrule height#2pt
        \hbox{\vrule width#2pt height#1pt \kern#1pt
              \vrule width#2pt}
              \hrule height#2pt}

\def\Fund#1#2{\vcenter{\vbox{\drawbox{#1}{#2}}}}
\def\Asym#1#2{\vcenter{\vbox{\drawbox{#1}{#2}
              \kern-#2pt       
              \drawbox{#1}{#2}}}}

\def\fund{\Fund{6.4}{0.3}}

\def\bfund{\overline{\fund}}

\newcommand{\Dsl}{\not\!\!D}
\newcommand{\be}{\begin{eqnarray}}
\newcommand{\ee}{\end{eqnarray}}

\begin{document}
 
\title{New Solutions to the Strong CP Problem}

\author{S.D.H. {\sc Hsu}}
 \email{hsu@duende.uoregon.edu}
 \affiliation{Institute of Theoretical Science, University of Oregon,
 Eugene OR 97403-5203}
\author{F. {\sc Sannino}}
 \email{francesco.sannino@nbi.dk}
\affiliation{NORDITA,  Blegdamsvej
  17, DK-2100 Copenhagen \O, Denmark\footnote{Since October 1st the address is: The Niels Bohr Institute, Blegdamsvej 17, DK-2100 Copenhagen \O, Denmark.} }

\begin{abstract}
We exhibit a solution to the strong CP problem in which
ultraviolet physics renders the QCD $\theta$ angle physically
unobservable. Our models involve new strong interactions beyond
QCD and particles charged under both the new interactions and
ordinary color.
\end{abstract}

\maketitle

\section{Introduction}

The CP violating QCD $\theta$ angle is the most mysterious of the
fundamental parameters of the standard model \cite{review}. $\theta$
is highly constrained by measurements of the neutron electric dipole
moment (NEDM): $\theta \lesssim 10^{-10}$. (We work in a basis where
${\rm arg~det} \, M_q = 0$, so that $\theta$ is directly related to
CP violation and the NEDM.) Theorists have long sought a mechanism
to explain why $\theta$ is so close to zero.

In this paper we exhibit a model in which new physics, at possibly
very high energy scales, renders $\theta$ an unobservable
parameter. The main idea is quite simple. We introduce a massless
fermionic field $Q$ which carries both ordinary color and also
transforms under another gauge group $SU(N)$. The anomaly
associated with the axial transformation of $Q$ receives a
contribution from gluons, so that phase rotations of $Q$ are
equivalent to shifts of the $\theta$ angle. Because $Q$ is
massless, there is a symmetry associated with these phase
rotations, allowing us to eliminate $\theta$ entirely. The $SU(N)$
interactions are necessary to bind $Q$ particles into heavy bound
states. These are not observed in low-energy physics, as there is
a mass gap which grows with $\Lambda_N$, the strong coupling scale
of the $SU(N)$ interaction. A novel aspect of our model is that
ordinary QCD is embedded in a left-right symmetric group $SU(3)_L
\times SU(3)_R$, where here 3 is not due to the number of light
quark flavors, but rather the number of colors in QCD. This
structure requires a non-standard Higgs sector to generate quark
masses as well as another fermion $T$ which is a color singlet and
charged only under the new $SU(N)$ group.

There are other viable solutions to the strong CP problem
\cite{Creutz,{Creutz:2004fi}} such as axion models, left-right
symmetric models and the ones in which CP is broken spontaneously
\cite{BN}. In axion models
\cite{PQ,Weinberg,Wilczek,Kim,SVZ,Dine:1981rt,Mohapatra:1978fy},
additional particles (usually heavy scalars and colored fermions)
are added to realize an anomalous Peccei-Quinn symmetry. The
observed value of $\theta$ QCD is then determined dynamically by the
location of the minimum of the instanton-generated potential for the
axion field, which is a pseudo-Goldstone boson of the Peccei-Quinn
symmetry. Quantum gravity is not believed to exhibit any {\it exact}
global symmetries, and even very small, Planck-suppressed,
violations of the Peccei-Quinn symmetry will spoil the axion
solution of the strong CP problem \cite{Planck}. In string theory,
it is possible to obtain global symmetries which are exact up to
violation by instanton effects. In this sense the axion solution of
the strong CP problem is generally considered to be very natural in
string theory (see for example \cite{Green:1987sp}).  However, the
axion mass scale, which is determined by a compactification scale,
is generally very close to the four-dimensional Planck scale. This
conflicts with some standard cosmological arguments which say that
such a scale should be no bigger than about $10^{13}$ GeV.

In our model we are left with an exactly massless Goldstone boson
\cite{Witten}. This Goldstone boson is not an axion (which would
have a non-zero mass), and our solution is not the standard axion
solution of the strong CP problem. A simple way to understand why
we are left with an exactly massless scalar field is to recall
that the new colored and uncolored fermions $Q$ and $T$ have zero
mass. This \cite{Bardeen-Tye,{Kandaswamy:1978wi}} prevents the
Goldstone boson from acquiring a mass term. Unlike in axion
models, the symmetry breaking scale associated with the Goldstone
boson can be taken to be arbitrarily large, effectively decoupling
it from ordinary particles.

\section{Model}

The model we consider is described in the accompanying table and
figure. (Generalizations are straightforward.) It contains the
symmetry groups $SU(N) \times SU(3)_L \times SU(3)_R$ and the
additional particles $Q$ and $T$, which are in the fundamental and
adjoint representation of $SU(N)$, respectively. The $SU(3)_c$ color
symmetry of QCD is the vector subgroup $SU(3)_V$ of $SU(3)_L \times
SU(3)_R$, so that QCD gauge transformations correspond to
simultaneous transformations with $U_L = U_R$. Each of $SU(N)$ and
$SU(3)_{L,R}$ (or equivalently, $SU(3)_{A,V}$) are gauged.
Additional $U(1)$ axial charges and symmetries are listed. The
anomaly-free linear combination, denoted $U(1)_{TAF}$ is gauged.
Note that in the table we have suppressed the standard model flavor
index $f=1,...,F=6$ (i.e., $f=$ up, down, strange, ... top).
\begin{table*}[H]
\begin{tabular}{c||cccccccc }
 & $SU(N)$ & $SU(3)_L$& $SU(3)_R$& $U_V(1)$&$U(1)_{TAF}$&$U(1)_T $&$U(1)_{QA}$&$U(1)_A$  \\
  \hline \hline \\
$T_C^C$& {\rm Adj} & $1$& $1$& $0 $&$1$&$1$&$0$&$0$  \\
  \\
  $Q^C_{c}$& $\fund$ & $\bfund$& $1$& $1$&$-\frac{N}{3}$&$0$& $1$& $0$  \\
  \\
  $\widetilde{Q}^c_{C}$& $\bfund$ & $1$& $\fund$&$-1$ &$-\frac{N}{3}$&$0$&$1$&$0$  \\
  \hline\\
 $q^{c}$& $1$&$\fund$ &$1$ &$-\frac{N}{F}$ &$\frac{N^2}{3F}$ &$0$ &$0$&$1$  \\
 \\
 $\widetilde{q}_{c}$& $1$&$1$ & $\bfund$&$+\frac{N}{F}$ &$\frac{N^2}{3F}$&$0$&$0$&$1$  \\
\end{tabular}
\caption{The table summarizes the symmetries and particle content of
the model. The usual quarks $q$ of QCD also have a standard model
flavor index $f=1,\ldots,F=6$ which we suppressed. All fields are
Weyl spinors. There are five independent global $U(1)$'s associated
to each Weyl fermion transformation. We can make two independent
anomaly free combinations, labeled by $U(1)_V$ and $U(1)_{TAF}$ in
the table while there are still three anomalous $U(1)$
transformations which are the remaining $U(1)$ transformations in
the table. The gauge group is $SU(N)\times SU(3)_L\times
SU(3)_R\times U(1)_{TAF}$.} \label{symmetric}
\end{table*}

Gauging $SU(3)_A$, as well as $U(1)_{TAF}$, precludes the usual
standard model masses for the quarks $q$. These have the form
(suppressing all indices except those of $SU(3)_L \times SU(3)_R$):
\begin{equation}
\label{qmass} {\cal L}_{\rm quark~ mass} = m_q  q^{c}
\widetilde{q}_c ~+~ h.c.~,
\end{equation}
which under a general $SU(3)_L \times SU(3)_R$ transformation
becomes
\begin{equation}
m_q q U_L \, U_R^\dagger \widetilde{q} ~+~ h.c.
\end{equation}
The quark mass (\ref{qmass}) is invariant under $SU(3)_c =
SU(3)_V$, transformations, which have $U_L = U_R$, but not under
$SU(3)_A$. Hence, the usual Higgs coupling to quarks is forbidden
by gauge symmetry; the interaction which gives masses to quarks
must involve additional particles charged under both $SU(3)_L$ and
$SU(3)_R$.

To overcome this restriction, we postulate a field \cite{reader} $H$ 
which transforms as $(\bar{3},3)$ under $SU(3)_L \times SU(3)_R$, carries
$U(1)_{TAF}$ charge $-2F/3$ ($F=N$ to insure gauge anomaly cancellations),
and is a singlet under electroweak symmetries. $H$ couples to quarks 
via the higher dimension operator \cite{reader}
\begin{equation}
\label{H2} H \, \phi \, q \widetilde{q}~~~,
\end{equation}
where $\phi$ is the usual Higgs boson \footnote{Another option is to 
introduce a scalar field $H$ which transforms
as $(\bar{3},3)$ under $SU(3)_L \times SU(3)_R$, carries
$U(1)_{TAF}$ charge $-2F/3$ ($F=N$ to insure
gauge anomaly cancellations), and has the electroweak charges of the
ordinary Higgs (i.e., it is a doublet under $SU(2)_L$). Then, quark
masses result from an interaction of the form 
$ H \, q \widetilde{q}$. In this case 
$\langle H \rangle$ must be of order the electroweak scale
\cite{Barr:1985ig}, which is probably ruled out since the
Goldstone boson would be detectable}.  The
potential for $H$ must be chosen so that $H$ develops a vacuum
expectation value. The Goldstone boson mentioned in the
introduction is a linear combination of the phase of $H$ and that
of the quark condensate $q \widetilde{q}$. In this case the vev of $H$ 
 can be much larger than the electroweak scale since the breaking of electroweak symmetry is
solely due to the ordinary Higgs $\phi$. Note that we do not allow
interactions among $H$ and $Q\widetilde{Q}$ leading to a mass term
for $Q$. That is, we assume that the $U(1)_{QA}$ symmetry is only
violated by the anomaly (i.e., instanton effects). This is similar
to the assumption made about the Peccei-Quinn symmetry in axion
models.

The condensate $\langle \widetilde{Q} Q \rangle \sim \Lambda_N^3$
spontaneously breaks $SU(3)_L \times SU(3)_R \rightarrow SU(3)_V$,
where the unbroken subgroup is QCD. The axial subgroup $SU(3)_A$
is spontaneously broken, and the corresponding Goldstone bosons
are eaten, leading to massive $SU(3)_A$ gauge bosons. It is
dynamically preferred for the $H$ condensate to align with
$\langle \widetilde{Q} Q \rangle$ in the $SU(3)_L \times SU(3)_R$
internal space, since when they are not aligned color is broken
and gluons become massive. As reviewed in \cite{Peskin}, the
contribution to the vacuum energy from Higgsed gauge bosons is
larger than that from massless gauge bosons, which generically
leads to groundstates with maximal unbroken gauge symmetry.

We expect the colored excitations associated
with rotations of $\langle H \rangle$  relative to $\langle
\widetilde{Q} Q \rangle$ in the $SU(3)_L \times SU(3)_R$ to
have mass much larger than the weak or TeV scale since the $H$ dynamics is unrelated to
electroweak breaking. We also expect the $T$ field to condense and spontaneously break
$U(1)_{TAF}$, so that any states containing $T$ constituents have
masses at least of order $\Lambda_N$.

\subsection{Summary of the UV Theory and dynamics}
Before describing the low energy theory it is instructive to display explicitly the  
UV theory corresponding to the fields shown in the table. 
The fermion kinetic terms are given by:

\begin{eqnarray}
{\cal L}_{\rm FKT} &=&  \bar{Q}_{L}(i\Dsl )Q_L +\bar{Q}_{R}(i\Dsl )Q_R +\frac{1}{2}\bar{T}(i\Dsl )T 
\nonumber  \\  
&+&\bar{\mathcal{Q}}_L (i\Dsl ) {\mathcal{Q}}_L +\bar{u}_R (i\Dsl ){u}_R + \bar{d}_R (i\Dsl ){d}_R  \ ,
\end{eqnarray}
with $Q_L=Q$ and $Q_R = \bar{\tilde{Q}}$. $Q$ and $\tilde{Q}$ are left handed Weyl fermions. $\mathcal{Q}_L$ represents the electroweak quark doublet, while $q_L=q$ and  $q_R=\bar{\tilde{q}}$ in the notation of the table. The generation index for the ordinary quarks is not explicitly shown.  More explicitly, the kinetic terms for the new fermions are: 
\begin{eqnarray}
&&\bar{T} i\gamma^{\mu} \left(\partial_{\mu} - iB^A_{\mu}\hat{T}^A_{\rm Adj} -i{\cal A}_{\mu}\right) T ~, \\
&&\bar{Q}_{L/R} i \gamma^{\mu} \left(\partial_{\mu} - iA^a_{{L/R}\mu}T^a - iB^A_{\mu}\hat{T}^A \mp i\frac{N}{3}{\cal A}_{\mu} \right) Q_{L/R}. \nonumber
\end{eqnarray}
$A_{L/R}^a$ are the gauge bosons of the color left and color right gauge interactions with $a=1,\ldots 8$. $B^A$, with $A=1,\ldots N^2-1$, are the gauge bosons of the new strong interaction. ${\cal A}$ is the gauge boson of the anomaly free $U(1)_{TAF}$ gauge symmetry. For the ordinary quarks the kinetic terms are:
\begin{eqnarray}
&&\bar{\mathcal{Q}}_{L} i\gamma^{\mu} \left(\partial_{\mu} - iA^a_{{L}\mu}T^a+ i\frac{N^2}{3F}{\cal A}_{\mu} - iW^b_{\mu}\tau^b  \right) {\mathcal{Q}_{L}} \ ,\nonumber \\
&&\bar{q}_R i \gamma^{\mu} \left(\partial_{\mu} - iA^a_{R\mu}T^a  - i\frac{N^2}{3F}{\cal A}_{\mu}\right) q_R \ .
\end{eqnarray}
Here we have indicated for illustration the interactions with the electroweak gauge bosons $W^a$, but neglected the hypercharge gauge boson. $F$ is the total number of quark flavors. To this Lagrangian one has to add the kinetic term for the new gauge bosons and the modified Yukawa interactions which lead to masses for the ordinary fermions. As mentioned earlier we are now forced to introduce another complex scalar field transforming under the left and right color transformations: $H^c_{c^{\prime}}$. The resulting Yukawa interactions are:
\begin{eqnarray}
-\frac{\lambda_d}{M}\bar{\mathcal{Q}}^{c^{\prime}}_L\cdot\phi H^c_{c^{\prime}} d_{R, c} -  \frac{\lambda_u}{M} \epsilon^{\alpha\beta}\bar{\mathcal{Q}}^{c^{\prime}}_{L,\alpha}\phi^{\dagger}_{\beta} H^c_{c^{\prime}} d_{R, c} + {\rm h.c.} \ .
\end{eqnarray}
$M$ is a scale related to the $H$ sector of the theory while $\phi_{\alpha}$ is the standard electroweak doublet $\alpha=1,2$. 
Other nicer ways of providing mass to the ordinary quarks can, of course, be explored - a more general structure of the Yukawa couplings in flavor space might be expected. With this choice of the Yukawa sector the leptonic sector of the standard model remains unmodified. It should also be clear that $H$ must condense for the ordinary quarks to acquire a mass.

The $SU(N)$ gauge theory, being vector like, is free from gauge
anomalies. Since we independently gauge $SU_{L,R}(3)$ we also need
to cancel the associated gauge anomalies. The simplest way is to
construct a vector like theory with respect to each gauge group.
This can be easily achieved by setting $N=6$, i.e. equal to the
number of ordinary quark flavors. For $N=6$ (recalling the 6
flavors of quarks $q_f$) we see that in each vertical column of
the table corresponding to a non-Abelian group the particle
content is vector like.

As for the summary of the dynamics we recall that our model has four independent scales. The scale of the $SU(N)$ strong dynamics $\Lambda_N$, the scale $M$ of the condensation of $H$, the electroweak scale and finally the ordinary QCD confining scale. We imagine the first two scales to be much larger than the electroweak scale.
Below the $SU(N)$ confining scale, as explained in the previous section, $SU(3)_L\times SU(3)_R \times U(1)_{TAF}$ breaks spontaneously to $SU(3)_V$. We identify $SU(3)_V$ with ordinary color interactions. The effective low energy theory below the $\Lambda_N$ scale - but remaining above the electroweak scale - is obtained by replacing the $SU(N)$ UV theory with its low energy chiral perturbation theory. The Goldstone modes (massless colored pions) become longitudinal components of the axial vector bosons and hence, in the end, disappear from the low energy theory. A similar fate is shared by the Goldstone boson associated to the gauged exact $U(1)_{TAF}$ symmetry. Since $T,Q,\widetilde{Q}$ are massless we have no $SU(N)$ theta term. By construction $Q$ carries ordinary color and we stress that, without invoking any chiral rotation of the ordinary quarks, the QCD $\theta$ term becomes unphysical (see next sections for a formal proof). 
This is the main point of our paper. 

At the electroweak scale, having already assumed the condensation of the field $H$ at some energy $M$ less than or of order the confining scale of $SU(N)$, one generates a mass term for the standard fermions via Yuakawa interactions. 
In general the couplings are complex and one might be worried that they regenerate a strong CP phase at low energies. Although the formal proof is provided in a following section we can immediately argue that such a strong CP phase cannot appear. This is {\it independent} of the mechanism we use to give masses to the ordinary quarks. To demonstrate this we can first perform a non-abelian chiral rotation which brings all the quarks to the same complex phase. Then we are left with an axial transformation which, due to the quark axial anomaly, potentially leads to a new strong CP phase. However we are free to perform an equivalent axial transformation of the $Q$ quarks to offset the strong CP phase again. 

Unfortunately, we were forced to introduce another field $H$ which also carries a new phase. In the absence of the $T$ and $Q$ fields this would lead to a conventional axion field solution to the strong CP problem. However as we shall demonstrate below the would-be axion in our model is exactly massless. It is important to note that our mechanism for rotating away the strong CP phase does {\it not} require this extra $H$ field and the hope is that a model similar to ours can be constructed in which such a field is not needed.

\subsection{The Massless Goldstone Boson}

We now construct the low energy effective theory for the 
pseudoscalar particles associated with the axial $U(1)$ symmetries in our model. 
At energies above the electroweak
scale we have the two condensates $\langle TT \rangle$ and
$\langle Q\widetilde{Q} \rangle$. We hence expect two independent
pseudoscalars, one from each condensate, which are 
singlets of the $SU(3)_L\times SU(3)_R$ non-Abelian
symmetries:  
\begin{eqnarray}
<TT> &=&|<TT>| e^{i\eta_T} \ , \nonumber \\
{\rm det}<Q\widetilde{Q}> &= &|{\rm det} <Q\widetilde{Q}>| e^{i\eta_Q} \ , 
\end{eqnarray}
by $\eta_T$ we denote $\eta_T/F_T$ and $\eta_Q$ is $\eta_Q/F_Q$. The decay constants 
$F_T$ and $F_Q$ are comparable in size and of the order of the confining scale of the $SU(N)$ gauge theory. 
Since the confining scale of $SU(N)$ is much larger than the electroweak scale the massive scalar excitations
will not appear in the low-energy theory. 
The instanton-induced effective potential, which preserves the $U(1)_{TAF}$ symmetry, is
\begin{eqnarray}
V_{TAF}= c_1 |<TT>|^N \,|{\rm det} <Q\widetilde{Q}>|e^{i(N\eta_T + 3\eta_Q)} +{\rm h.c.}  \ , 
\end{eqnarray} 
with $c_1$ a constant. This potential determines the linear combination of pseudoscalars which becomes massive. The orthogonal combination remains massless. It is absorbed in the longitudinal component of the $U(1)_{TAF}$ gauge boson and hence decouples from the low energy physics. The $SU(N)$ $\theta$ angle is rendered an unphysical
quantity since $T,Q,\widetilde{Q}$ are exactly massless. 

At much lower energies we include the effects of the ordinary quarks. To
give mass to the fermions while preserving the $U(1)_{TAF}$ symmetry 
we introduced the field $H$. Note that, even in the presence of a
mass term for the quarks, the QCD theta angle is unobservable, since
the massless $T,Q,\widetilde{Q}$ fields allow us to rotate away {\it
both} the $SU(N)$ and QCD theta angles. (We discuss this issue
further in the following subsections.) The phase of $H$ combines
with the phase of $q \widetilde{q}$ yielding a massless Goldstone
boson and a massive pseudo-Goldstone boson. The latter is the
$\eta^{\prime}$ meson of QCD. To show this one can use either
current algebra
 techniques \cite{Bardeen-Tye} or the effective Lagrangian approach
\cite{Kandaswamy:1978wi}. 
We present here the effective Lagrangian
approach. The two new pseudoscalars are related to the phase 
$h$ of $<H>$ and $\eta_q$ of $<q\widetilde{q}>$. 
At very low energies, below the QCD scale, we deduce the following
effective potential, where we restrict our attention to pseudoscalar fields (recall
$F = 6$ is the number of quark flavors, not a decay constant):
\begin{eqnarray}
V_{\rm Low} &=& c_2 |<q H\tilde{q}>|e^{i(h+\eta_q)} \nonumber \\ 
&+& c_3|<{\rm det}\left[qH\tilde{q}\right]>|e^{iF(h+\eta_q)} + {\rm h.c.} \ . 
\end{eqnarray}
The last term is the ordinary instanton-induced fermion determinant. 
It is clear that the only combination which acquires a mass is $h+\eta_q$. 
Here $h$ stands for $h/|<H>|$ and  $\eta_q$  for $\eta_q/F_{\pi}$.  The 
linear combination shown in the potential is the standard 
$\eta^{\prime}$ of QCD, while the orthogonal combination 
$h - \eta_q$ remains massless. 

We stress that $h - \eta_q$ is a true massless Goldstone boson, 
not an axion of the
usual type. While the QCD $\theta$ angle has become {\it
dynamical}, in the form of the massless linear combination, the
presence of the massless fermions $Q,\widetilde{Q}$ and $T$ has rendered physics
independent of $\theta$, and hence the potential for the massless
combination flat. The result is similar to that of an axion model
that {\it also} has a massless quark, except in this case it is
the $Q$ and $T$ fields which play the role of the massless quark.
In usual axion models, two different linear combinations of the axion field and 
$\eta_q$ appear: one induced by the mass term of the quarks and 
the other due to instantons \cite{Kandaswamy:1978wi}. This leads to both
a massive axion and a massive $\eta'$, unlike in our model.

\begin{figure}[h]
\begin{center}
\includegraphics[width=7truecm,height=6truecm,clip=true]{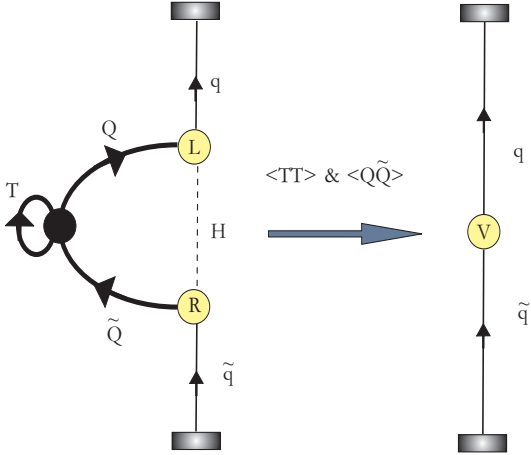}
\caption{Schematic representation of the model. The black node
represents the $SU(N)$ gauge theory. The left (right) node indicates
$SU(3)_{L(R)}$.  After condensation of the $T$  and $Q$ fermions the
$V$ node is the QCD gauge group. The two squares represents the
ungauged $SU(F)$ left and right QCD flavor groups.} \label{figure}
\end{center}
\end{figure}

\subsection{Axial Anomalies}
We have three axial currents, and only one is anomaly free with our
charge assignments:
\begin{eqnarray}
\label{AWI}
\partial_{\mu}J_{TAF}^{\mu} &=& 0
\ , \nonumber \\
\partial_{\mu}J_{QA}^{\mu} &=& \frac{1}{16\pi^2}\left[3 F_{\mu\nu}\widetilde{F}^{\mu\nu} +
N G_{\mu\nu}\widetilde{G}^{\mu\nu} \right] \nonumber \\
\partial_{\mu}J_{A}^{\mu} &=& \frac{1}{16\pi^2}\left[6 G_{\mu\nu}\widetilde{G}^{\mu\nu}
\right] \ .
\end{eqnarray}
Here we have denoted with $F^{\mu\nu}$ the $SU(N)$ field strength
while the associated gauge field is $B^{\mu}$. In the anomaly
equations $G_{\mu\nu}\widetilde{G}^{\mu\nu}$ represents
\begin{eqnarray}
G_{\mu\nu} \widetilde{G}^{\mu\nu} = \frac{1}{2} \left(
{G_{\mu\nu}}_L \widetilde{G}^{\mu\nu}_L + {G_{\mu\nu}}_R
\widetilde{G}^{\mu\nu}_R \right) \ ,
\end{eqnarray}
where $G^{\mu\nu}_{L,R}$ are the field strengths for the
$SU(3)_{L,R}$ gauge groups, whose gauge fields are
$A_{L,R}^{\mu}$.

At scales much below the $SU(N)$ confining scale $\Lambda_N$ the
axial gauge bosons can be integrated out. In this limit the field
strength $G_{\mu \nu}$ appearing on the rhs of equations
(\ref{AWI}) is just the usual QCD gluon field strength.

The anomalous Ward identities (\ref{AWI}) imply that axial
rotations of the $T$ field are equivalent to shifts of the $SU(N)$
$\theta$ angle (henceforth denoted $\theta'$). Axial rotations of
$Q$ shift $\theta'$ as well as the $SU(3)_{L,R}$ angles denoted
$\theta_{L,R}$, while $q$ rotations only shift $\theta_{L,R}$. It
is clear that we can eliminate $\theta'$ by appropriate $T$
rotation. $\theta_{L,R}$ are discussed below.

\subsection{Euclidean functional integral}

The partition function of our model is
\begin{eqnarray} Z &=& \sum_{\mu \nu_L \nu_R}~
\int \left[ DB \right]_\mu \left[ DA_L \right]_\nu \left[ DA_R
\right]_{\nu'} ~e^{-S(A_L, A_R, B)} \nonumber
 \\&& \times ~{\rm det} \, Q ~ {\rm det} \, T~ {\rm det} \,
q~ e^{i \mu \theta^{\prime} + i \nu_L \theta_L  + i \nu_R \theta_R
}~~,
\end{eqnarray}
where the fermionic integrals have been performed leaving
determinants of the respective Dirac operators.

The measure of the integral has been divided into winding number
sectors, where the $SU(3)_{L,R}$ winding numbers are given by
$\nu_{L,R} = \frac{1}{16\pi^2} \int d^4x~ {G_{\mu \nu}}_{L,R}
\tilde{G}^{\mu \nu}_{L,R} (x)$, and the $SU(N)$ winding number is
$\mu = \frac{1}{16\pi^2} \int d^4x~ F_{\mu \nu} \tilde{F}^{\mu
\nu} (x)$. It is convenient to define:
\begin{eqnarray}
\nu & = & \frac{1}{2}(\nu_L + \nu_R) \ , \qquad
\nu_A=\frac{1}{2}(\nu_L -
\nu_R) \ , \\
\theta &=& \theta_L + \theta_R \ , \qquad \theta_A = \theta_L -
\theta_R \ .
\end{eqnarray}
The left-right symmetry of our model suggests that $\nu_A=0$
implying that the physics is insensitive to $\theta_A$. To arrive
at the same conclusion we can use the following dynamical
argument. Due to the Higgs mechanism, the axial gauge fields
$A_L^{\mu}-A_R^{\mu}$ are heavy. At low energies their
fluctuations are suppressed, or equivalently: $A_L^{\mu}=
A_R^{\mu}$. Consequently, $\nu_L= \nu_R$, which implies that
$\theta_A$ is an unphysical parameter.

We can then concentrate on the remaining angles: $\theta^{\prime}$
and the usual QCD $\theta$. Here we can use the index theorem
relating the number of chiral zero modes of the $T$ Dirac operator
to the Pontryagin index:
\begin{equation} n^T_+ - n^T_- = N \mu~~.
\end{equation}
For the $Q$ operator, we have
\begin{equation}
n^Q_+ - n^Q_- = 3 \mu + N \nu~~,
\end{equation}
and for the $q$ operator
\begin{equation}
n^q_+ - n^q_- = 6 \nu ~~.
\end{equation}
Here $n_\pm$ denotes the number of zero modes of chirality $\pm$.
Because $Q$ and $T$ are massless, their determinants vanish
whenever there are any zero modes - in other words, if {\it any} of
the respective $n_\pm$ are non-zero. The condition of non-vanishing
determinants requires $\mu = 0$ and $\nu = 0$. The only topological
sectors that can contribute have $\mu = \nu = 0$. This means that
the partition function is independent of both $\theta$ and $\theta'$
- they are unphysical parameters.

Note also that it is sufficient that $T$ and $Q$ are massless to
completely rotate away $\theta$ and $\theta^{\prime}$. This is
important since it allows, at low energies, the quarks to acquire
a mass term without upsetting our results.

\bigskip
\section{Conclusions}

We described a class of models in which the QCD $\theta$ angle is
rendered unobservable by new short distance physics involving a new
strong force $SU(N)$. A generic prediction of the models is new
colored particles and a non-standard Higgs structure for the quark
masses. Interestingly we have an exactly massless pseudoscalar boson
which may be relevant for cosmology and is similar to a Majoron, for
which constraints from accelerators and astrophysics have been
recently analyzed \cite{Hannestad:2002ff}.


\bigskip
\subsection{Acknowledgements}
We thank H. Georgi, D.K. Hong, P. Olesen, M. Shifman,
K. Tuominen, M. Wise and E. Witten for discussions and useful
comments. G. Gabadadze is thanked for discussions concerning a
similar model he presented at the Workshop on Continuous Advances
in QCD 2004, Minneapolis, Minnesota, 13-16 May 2004. S.H. thanks
NORDITA and the Niels Bohr Institute for their hospitality while
this work was initiated. The work of S.H. was supported in part
under DOE contract DE-FG06-85ER40224.



\begin{thebibliography}{9}
\bibitem{reader} The alert reader might wonder why we do not use
$\widetilde{Q} Q$ in place of $H$ in equation (\ref{H2}). The
reason is that the $Q$ phase rotation used to eliminate $\theta$
QCD would then re-introduce $\theta$ into the quark mass matrix.
This is also why the $U(1)_{QA}$ symmetry must be violated only by
the anomaly and not by other explicit interactions.
\end{thebibliography}
\end{document}